# Volatility and irregularity Capturing in stock price indices using time series Generative adversarial networks (TimeGAN)

Leonard Mushunje[1], David Allen[2] and Shelton Peiris[3]

[1]Columbia University, Department of Statistics

[2,3]University of Sydney, School of Mathematics and Statistics

**Abstract**

This paper captures irregularities in financial time series data, particularly stock prices, in the presence of COVID-19 shock. We conjectured that jumps and irregularities are embedded in stock data due to the pandemic shock, which brings forth irregular trends in the time series data. We put forward that efficient and robust forecasting methods are needed to predict stock closing prices in the presence of the pandemic shock. This piece of information is helpful to investors as far as confidence risk and return boost are concerned. Generative adversarial networks of a time series nature are used to provide new ways of modeling and learning the proper and suitable distribution for the financial time series data under complex setups. Ideally, these traditional models are liable to producing high forecasting errors, and they need to be more robust to capture dependency structures and other stylized facts like volatility in stock markets. The TimeGAN model is used, effectively dealing with this risk of poor forecasts. Using the DAX stock index from January 2010 to November 2022, we trained the LSTM, GRU, WGAN, and TimeGAN models as benchmarks and forecasting errors were noted, and our TimeGAN outperformed them all as indicated by a small forecasting error.

## Background and Introduction

In financial markets, the global crisis of COVID-19 is nearly analogous to the effects of the Global financial crisis of 2007/2008, Mushunje, (2021). The pandemic has significantly changed the investment styles in all markets, including Capital and Money markets. On the same note, modeling any related COVID-19 shocks with univariate and multivariate time series models such as ARCH, GARCH-VAR, and single equation models has been tremendously affected, Bobeica and Hartwig, (2021). Investigating the impact of COVID-19 observations on forecasting stock prices is a challenging path of interest. Stabilizing the parameters after adding the COVID-19 observations also stabilizes the unconditional forecast paths, and this is not easy when using models like the VAR.

Modeling shock effects in markets is complex due to the propelled stylized statistical facts such as kurtosis, commonly called tail risk. This tail risk is associated with the widely known risk measure-VaR. Tail risk implies the realization of high probabilities relative to the normal distribution, Mishkin, (2011). Since the inception of the pandemic in 2019, macroeconomists, quants, and other practitioners have needed help to make sense of their existing models (inputs and outputs). This was because the pandemic triggered some shocks that brought instability to the models and abruptly changed the internal modeling cultures. In some markets, such as labor markets, the effects of the pandemic were characterized by colossal time series variations, for example, actual activities and labor market key pointers. For more, see, (Schorfheide and Song (2020), Lenza and Primiceri (2020), and Carriero et al. (2021)). Given all such instances, we provide a hybrid alternative to shock capturing when dealing with sequential data- Financial Time series data.

Capturing financial data irregularities and volatility risks is critical to market traders and investors, especially during COVID-19. Therefore, analysts and practitioners must have models capable of capturing these facts and providing valid and meaningful results before advancing to any critical decisions. However, building such models requires dynamic and hybrid statistical methods. This paper proposes one of them, which is based on data and distribution learning (Time GANS). This model can provide useful information about several complex stylized facts, fat_tails, long-range dependence, and kurtosis, see Shuntaro et al. (2019). Stylized facts and price oscillations have been long modeled using physics-inspired models, for example, the Brownian motion (Heston, (1993) and the Heston model (Mushunje, 2019).

We see salient work on time series modeling and Generative adversarial networks with applications to finance. Two major approaches that are common in complex financial time series modeling include stochastic processes and agent-based modeling (Bollerslev, T. (1982), Engle, R. (1986), Chalet, D., Zhang, Y-C. (1999)). Stochastic nature models are based on temporal-dependent parameters, including ARCH and the GARCH. Whereas agent-based modeling primarily describes how different agents interact with each other in the market, given any set of parameters. Nevertheless, these models are not superior regarding stylized fact capturing (irregularities, kurtosis, and so on) is concerned, especially in the presence of shocks. Generative adversarial networks are featured in financial modeling as an advancement and extension of the already existing time series models, referred to as traditional time series models in this paper. As previously mentioned, we have a broad family of time series models, including the discrete-time ARCH-GARCH models (Bollerslev, 1986), as well as continuous time models like the most known Black and Scholes (1973) and Heston (1993). Nevertheless, extending these models takes work. For example, the Black Scholes model was extended to the Heston type, which uses stochastic volatility after nearly 20 years of deep and continuous research. This reflects the need for more financial innovation despite its deep and severe need. Machine learning and deep modeling in finance are exciting focus points, especially with the beauty of high-performing machines and other tools like graphical processing units (GPUs). We, therefore, summarize the innovations related to our current work on GANS as follows. Koshiyama et al. (2019) proposed a GANs model with a conditional variant and parameter scoring feature, and Pardo (2019) looks at the Wasserstein natured GANs.

Volatility clustering with GANs is seen in Jolicoeur-Martineau (2018), in which they additionally described the value derived from synthetic paths in volatility analysis and strategic trading. This is also found in Pardo and Lopez (2019). The idea of volatility clusters closely relates to the work of Takahashi et al. (2019) on the application of GANs in computing stylized facts embedded in stock features. Recurrent approaches to GANs were found helpful by Wiese et al. (2019) (Buehler & Ryskin, 2017). Given all these instances, deep learning, especially deep generative models, is a promising solution to these modeling g challenges. Some notable contributions to GANS can be seen in Radford et al. (2015) and Donahue et al. (2018). In financial time series modeling, we have works by Zhou et al., 2018 Fiore et al.,
2019; Koshiyama et al, 2019). This paper uses deep learning approaches to fill the gap in stylized fact capturing and modeling in financial time series data in times of COVID-19 shock. In a nutshell, our main contributions to this study are summarized as follows:

(i). First, we apply the GANs to explore and investigate the correlation and causal effects between the stock index attributes (prices, returns, volumes) and COVID-19 events such as case announcements and lockdown news announcements. The idea is to investigate if the pandemic affects stock market performance. Thus, we shall use pre and post-COVID-19 stock index data. (ii). We aim to assess the relationship between real-time COVID-19 news announcements and real-time stock variable movements in selected stock indices using GANs. The idea is to efficiently model stochastic irregularities and associated volatility structures through temporal dependence capturing. This enhances a real-time stock market intraday trading performance prediction. This aids investors in decision-making by stimulating efficient portfolio mixing. (iii). There are unpredictable and unexpected losses-fat tail effects in stock markets behind the COVID-19 pandemic, which needs attention. Thus, modeling the time series of the stock market data using GANs before and after COVID-19 aids in identifying the underlying these rare but devastating stylized facts. Here, we present a one-to-one idea mapping of COVID-19 and the 2008 global financial crisis. We aim to apply the GANs to financial time series stock data to uncover these stylized facts. This facilitates supernormal returns by minimizing potential (expected and unexpected) losses. Some underlying targeted facts are the fat-tailed price return distribution, volatility clustering, and leverage effects.

**Traditional Time Series Models**

Traditional time series frameworks such as ARCH/GARCH family models rely on classical statistics. These models measure the change in variance over time in a time series by capturing the variance of the current error term as a function of previous errors. Ideally, the relationship is between past innovations' squares and current errors' variance. Auto-regressive (AR) models predict future variances from historical information. They are commonly used in time series modeling where time-varying volatility and volatility clustering are present. However, these models lack a random component since volatility depends on historical values. On the other hand, ARCH/GARCH models predict the variances.

Another class of time series models is the Agent-based family (ABM). In Agent-based models, objects are mimicked by entities (agents). States are represented by agents that take any form of data. The primal goal behind ABMs is the multiple simulations of interactions that re-create and predict the behavior of complex phenomena. Modeling financial time series is a big challenge since the dynamics of financial markets are very complex, and the mechanism generates the data. Thus, its original distribution still needs to be discovered. Adopting a purely data-driven modeling approach to this problem could provide new solutions or alternative paths by removing a source of bias from modeling. In this study, we apply time series Generative Adversarial Networks (TimeGANS) to generate and predict stock prices to capture and identify any embedded stylized facts, such as volatility clusters and stochastic jumps. GANs provide good results since they can generate data by sampling only from real data, often with no additional assumptions or inputs. Avoiding assumptions may be an essential aspect of this type of modeling due to the largely empirical aspect of financial data. Avoiding possible human bias infiltrating the modeling process could be a step forward in this study.

## Methodology

### Data

The study employed the DAX stock index for a period spanning from January 2010 to November 2022, taking into account the weekend and holiday effect. Factoring out the non-trading days improved the tidiness of our data and reduced noise in our data. We detected some data that needed to be included, mainly due to non-trading effect and unavailability. To address this, we respectively extrapolated the data where the data from Friday is mapped to the following Monday and applied the Kth Nearest Neighbour approach. This data set represents the real set for our TimeGAN model, which will be used to generate the fake ones necessary for training. We provide below a full summary description of the variables in use.

### Variable explanation

| Variable | Description |
|---|---|
| Open Price | The stock price at the start of each trading day. |
| Close Price | The stock price at the end of each trading day. |
| High Price | The highest recorded trading price of each stock. |
| Low Price | The lowest recorded trading price of each stock. |
| Adj Close | The close price after considering all corporate actions (dividends, new offerings, and stock splits) |
| Volume | Number of stocks traded during a given period |
| Open_Diff | Differences in the open prices of stocks on consecutive days. |
| Close_Diff | Differences in the close prices of stocks on consecutive days. |
| Volume_Diff | Trading volume differences on consecutive days. |
| Adj Close_Diff | The difference between adjusted close prices at different consecutive days. |
| High_Diff | Differences in the highest prices reached by stocks on consecutive days. |
| Low_Diff | Differences in the lowest prices attained by stocks on consecutive days. |
| Open_SMA | Simple moving averages for open stock price over the period under study. |
| Close_SMA | Simple moving averages for close stock price over the period under study. |
| High_SMA | Simple moving averages for the highest prices reached over the period under study. |
| Low_SMA | Simple moving averages for the lowest prices reached over the period under study. |
| Adj Close_SMA | Adjusted Close price simple moving averages over the study period. |
| Volume_SMA | Simple moving averages for trading Volume over the period under study. |

*Table 1: We describe the variables used in this study. In addition to the main variables extracted from the DAX stock index, we created additional useful variables as given above. Firstly, the differences are computed as percentage changes between two consecutive stock prices and volumes. We also computed simple moving averages for all our features on a 10-day interval.*

**GAN Background**

**Generative Adversarial Networks (GANs)**

These are some of the deep generative models used to generate realistic data by training deep neural networks. GANs consist of two different neural networks, a generator *G* and a discriminator *D*. The generator *G* is responsible for the generation of data, and the discriminator *D* functions to judge the quality of the generated data and provide feedback to the generator *G*. These neural networks are optimized under game-theoretic conditions: the generator *G* is optimized to generate data that deceive the discriminator *D* and the discriminator *D* is optimized to distinguish the source of the input, namely the generator *G* or realistic dataset. Below, we briefly describe some primary forms of Financial time series GANS models. Unlike other time series models, GANs are maximum likelihood-free. We do not necessarily implement such techniques when training GANs and related.

**GAN Model Parameters**

| **Parameters** | **GAN generator** | **GAN discriminator** | **GAN model** |
|---|---|---|---|
| Learning rate | 0.00001 | 0.00001 | 0.000 1 |
| Batch size | 128 | 128 | 128 |
| Number of epochs | 100 | 100 | 250 |
| Optimizer | Adam | Adam | Adam |
| Loss function | Sigmoid | ReLU & Sigmoid | Cross-entropy |
| Dropout | 0.4 | | |
| Strides | | 4 | |
| Layer 1 | GRU units = 1024 | CNN lD = 32 | |
| Layer 2 | GRU units = 512 | CNN I D = 64 | |
| Layer 3 | GRU units = 256 | CNN lD = 128 | |
| Layer4<br><br>Layer 5 | Dense 128<br><br>Dense 64 | Flatten Dense 220 | |
| Layer 6 | Output | Dense 330 | |

*We provide a summary of the parameters used in our modeling framework. These parameter values are fixed throughout the modeling process. However, we initially changed the number of epochs and layers by cross-validation during training. The idea is to find an optimal set of parameters we displayed above.*

**Likelihood-free learning**

The natural questions are: What is likelihood-free, and why is it adopted in the GANs model family? Ideally, optimal generative models provide high sample qualities with the highest test log-likelihood. However, log-likelihood numbers tend to feature poor sample qualities as well. Given all this, we ignore likelihood estimation techniques and employ a likelihood-free training approach.

Naturally, we can formulate a likelihood-free objective function by considering two samples from distributions, say, $P$ $and$ $Q$, and applying a two-sample test to check if data from two samples follow the same distribution. Unambiguously**,** for two samples, $S_1$ $and$ $S_2$, such that $S_1 = \{\boldsymbol{X} \sim P\}$ $and$ $S_2 = \{\boldsymbol{X} \sim Q\}$, we can find a test statistic $M$, based on the difference between $S_1$ *and* $S_2$. The statistic is compared to the level of significance $\alpha = 5\%$, which we use to decide on the Null versus Alternative hypothesis that $P = Q$.

The same approach is used in our Generative adversarial network training setups. We have a training set/sample, $S_1 = \{\boldsymbol{X} \sim pdata\}$ and a test set, $S_2 = \{\boldsymbol{X} \sim p\theta\}$, where the underlying theme is to train the GAN model while minimizing the error encountered when training the two samples $S_1$ *and* $S_2$. However, the objective is not easy to train in high dimensional spaces, a common problem in GAN training. We shall devote to surrogate objective training in this study.

**GAN original Model Architecture**

In this section, we formulate the architecture for the GAN model. We have two nodes/networks gaming against each other in a competition setup. The generator, $G_\theta$, generates the sample $X$ from the original data set $Z$. The samples are generated deterministically and passed on to the other network, the discriminator, for classification. The discriminator $D\phi$ is a function that trains, distinguishes, and classifies the real and generated data. Below is the schematic form of the GAN architecture in question.

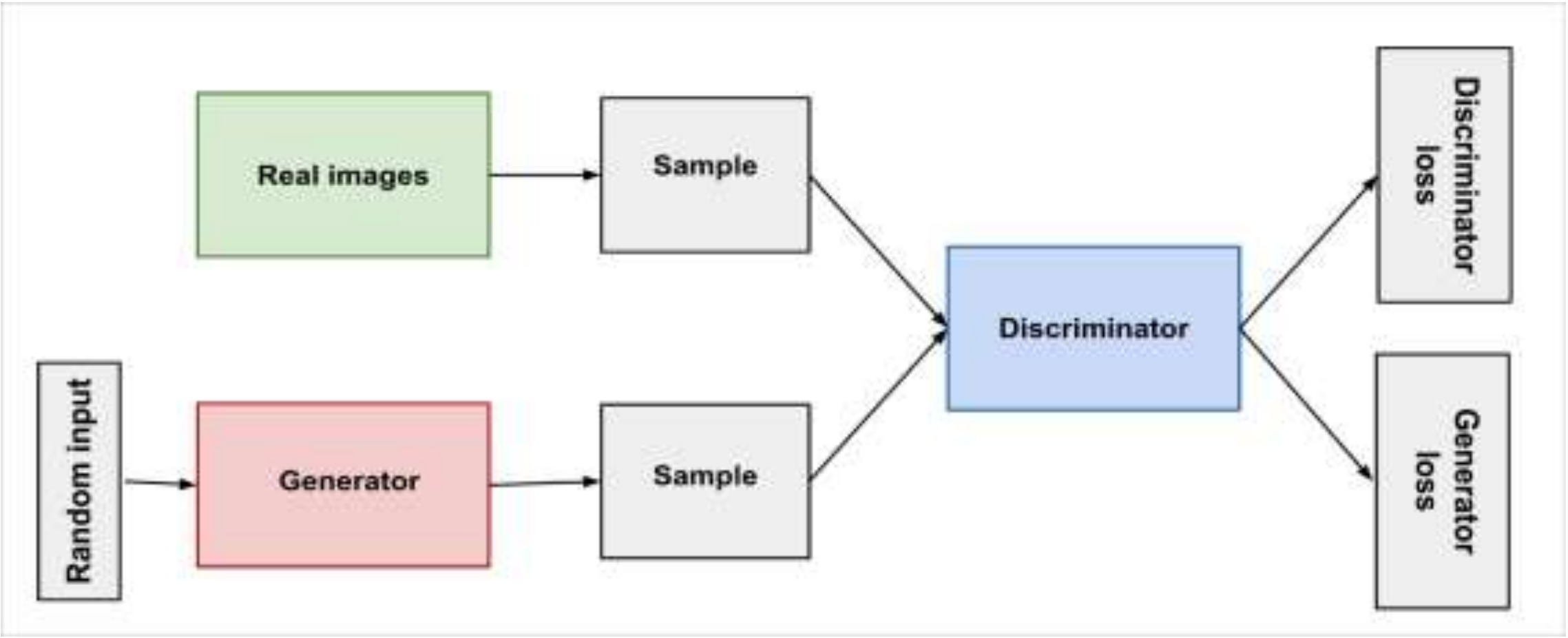

*Figure 1: GAN Architecture: The figure depicts all the states and output, including the connections between all the networks that comprise an entire GAN network. Firstly, the generator network takes in the random input before passing it on to the discriminator network. On the other hand, the real input is passed directly to the discriminator function. The discriminator then classifies the output as either real or fake. This comes with loss functions for each classification, as we shall further discuss.*

Our modeling approach is centered on the minimax game structure, where the generator attempts to minimize the objective of the form $p_{data} = p_{\theta}$, and the discriminator tries to maximize the function $(p_{data} \neq p_{\theta})$. In this setup, the discriminator is fooled by the generator by generating multiple samples that are difficult to differentiate from the real ones

The objective function under study is structurally presented as follows:

$$\min_{\theta}\max_{\phi}(V(G\theta, D\phi)) = \mathbb{E}[\boldsymbol{X}] \sim \boldsymbol{pX}[logD\phi(\boldsymbol{X})] + \mathbb{E}[\boldsymbol{Z}] \sim p(\boldsymbol{Z})[log(1 - D\phi(G_{\theta}(\boldsymbol{Z})))]$$

$\boldsymbol{Z}$ is the random input data, and $\boldsymbol{X}$ is the actual/real data from our learning model. During the training, the discriminator maximizes the objective function above with respect to $\phi$. Thus, for given samples generated by $G_{\theta}$, the discriminator classifies it by assigning a probability of one if $\boldsymbol{X} \sim p_{dat}$ and zero if $\boldsymbol{X} \sim p_{G}$. This yields the optimal setup as follows:

$$D * G(\boldsymbol{X}) = pdata(\boldsymbol{X})pdata(\boldsymbol{X}) + pG(\boldsymbol{X})$$

Analogously, the generator minimizes the objective function, fixing the discriminator $D\phi$. This eventually gives the optimal discriminator function, $D * G(\cdot)$, necessary for the sufficiently reduced overall objective function: $V(G_\theta, D * G(\boldsymbol{X}))$. This is further worked on to provide the distance measure based on the Jenson-Shannon Divergence (JSD) approach:

$$JSD = 2D_{JSD}[p_{data}, p_G] - log4$$

The *D*JSD satisfies all the properties of the KL divergence, a symmetric distance measure commonly in use. Using this distance measure, the GAN generator function turns out to be $p_G = p_{data}$, and the final objective function we get using optimal generators and discriminators becomes $D * G * (\boldsymbol{X})\ is\ - log4$.

**GAN Training Algorithm**

The steps making up our model and training are summarized as follows:

For epochs 1,…, *N*

do:

1. Sample minibatch of size *m*

from data: $\boldsymbol{X}(1), \ldots, \boldsymbol{X}(m) \sim D$

Sample minibatch of size *m* of noise: $\boldsymbol{Z}(1), \ldots, \boldsymbol{Z}(m) \sim p_Z$

Take a gradient *descent* step on the generator parameters *θ*:

▿ *θV* (*Gθ*, *Dϕ*) =1*m* ▿ *θ*∑*i*=1*m*log(1−*Dϕ*(*Gθ*(**z**(*i*))))

Then, take a gradient *ascent* step on the discriminator parameters *ϕ*:

▿ *ϕV* (*Gθ*, *Dϕ*) =1*m* ▿ *ϕ*∑*i*=1*m*[log*Dϕ*(**x**(*i*)) +log(1−*Dϕ*(*Gθ*(**z**(*i*))))]

These steps are critical in our analysis, and we also used these to train the WGANs extension of the GANs model.

GANS is used for generative modeling, and ideally, as shown above, the generator receives random input vectors, *z*, from the latent space. These are then mapped to the real data vectors through training and fine parameter tuning and selection. When dealing with time series data, like in our case, the input is provided as follows:

$$\begin{bmatrix} z_{11} & \cdots & z_{1,i} \\ z_{2,1} & z_{22} & z_{2,i} \\ z_{n,1} & z_{n-1,i-1} & z_{n.i} \end{bmatrix}$$

**Cost Functions**

Discriminator's cost function

When training GANS, the discriminator function typically uses the same costs function. The engineering effect only takes place on the generator's function. Thus, in this study, we used the same cost function for the discriminator function across all the different GANS architectures. We used the following cost function.

$$J^{(D)}\left(\theta^{(D)}, \theta^{(G)}\right) = -\frac{1}{2} E_{x\sim pdata} \log(D(x) - \frac{1}{2} E_z \log\left(1 - D(G(z))\right))$$

This function is minimized during training, and the discriminator uses the same optimal strategy. The only exception is that the classifier is trained on two different data sets: one from real data and the other from generated/fake data. It is from this training that we get the ratio estimate of the form

$\frac{p_{data}(x)}{p_{model}(x)}$, for all x.

This function allows us to obtain and compute the divergences and respective gradients. In any setup, this is the differential aspect between GANS models and other variational autoencoders and Boltzmann machines.

**Minimax**

The zero-sum game is one of the main game setups in practice where the sum of the player's costs is zero. On the other hand, the minimax game is highly amenable in theoretical terms. Goodfellow et al. (2014b) first implemented this version of the GAN minimax function to mimic the minimization of the Jensen-Shannon divergence using the real and model data. This also involves game convergence in the players' equilibrium if the player's policies are updated directly in the function space. The different functions used and the optimization argument are given below.

$J^{(G)} = -J^{(D)}$, Zero-sum game.

$V\left(\theta^{(D)}, \theta^{(G)}\right) = -J^{(D)}\left(\theta^{(D)}, \theta^{(G)}\right)$, value function with the discriminator's payoff.

${\theta^{(G)}}^{*} = arg \min_{\theta^{(G)}} \max_{\theta^{(D)}} V\left(\theta^{(D)}, \theta^{(G)}\right)$, outer-loop minimization, and inner-loop maximization function.

In this study, we implemented the Jensen-Shannon distance as indicated above. When training the models, the deep neural nets represent the players, and parameter space represents the updated ground.

**Wasserstein GAN Model**

Below is a summary of how the Wasserstein Generative Adversarial Network (WGAN) model is implemented.

**Require:** : $\alpha$, the learning rate. $c$, the clipping parameter. $m$, the batch size. $n_{\text{critic}}$, the number of iterations of the critic per generator iteration.
**Require:** : $w_0$, initial critic parameters. $\theta_0$, initial generator's parameters.
**while** $\theta$ has not converged **do**
  **for** $t = 0, ..., n_{\text{critic}}$ **do**
    Sample $\{x^{(i)}\}_{i=1}^{m} \sim \mathbb{P}_r$ a batch from the real data.
    Sample $\{z^{(i)}\}_{i=1}^{m} \sim p(z)$ a batch of prior samples.
    $g_w \leftarrow \nabla_w \left[\frac{1}{m}\sum_{i=1}^{m} f_w(x^{(i)}) - \frac{1}{m}\sum_{i=1}^{m} f_w(g_\theta(z^{(i)}))\right]$
    $w \leftarrow w + \alpha \cdot \text{RMSProp}(w, g_w)$
    $w \leftarrow \text{clip}(w, -c, c)$
  **end for**
  Sample $\{z^{(i)}\}_{i=1}^{m} \sim p(z)$ a batch of prior samples.
  $g_\theta \leftarrow -\nabla_\theta \frac{1}{m}\sum_{i=1}^{m} f_w(g_\theta(z^{(i)}))$
  $\theta \leftarrow \theta - \alpha \cdot \text{RMSProp}(\theta, g_\theta)$
**end while**

*Figure 2: WGAN Algorithm in a high-level arena.*

WGANS makes use of the notion called weight clipping and Lipschitz constraints. The clipping technique helps enforce the Lipschitz constraints, and if the clipping weight is high, the weights take time to converge to the targets, which is true otherwise. This makes it challenging to train the model optimally. Also, if the weight clip is small, there is a risk of vanishing gradients, especially when the number of layers is high unless batch normalization is used.

It is always desirable to derive an optimal training level with reliable gradients. The fact that the EM distance is continuous and differentiable implies that the neural nets can be trained critically and optimally. Analogously, the Wasserstein model can achieve the same results if trained multiple times, and the Wasserstein function is continuous and differentiable everywhere. WGAN is more potent at dealing with vanishing gradients than the original GANs. It only needs intelligent parameter tuning and selection with clean data.

**Long-term short memory (LTSM)**

The recurrent neural network (RNN) suffers from the problem of vanishing gradient problem. This, in turn, resulted in the development of the LSTM model. Thus, the LSTM is just an extension of the RNN model. LSTM is a sequential persistence, memory-based model. Unlike RNN, which does not remember long-term dependencies, the LSTM model captures this shortcoming by capturing both short and long-term dependencies. Thus, the LSTM can model and measure time series data in high dimensions over long periods.

**LSTM Architecture**

As shown above, the LSTM comprises three main gates: the Forget gate, Input gate**,** and Output gate**.** The LSTM model contains the hidden states designed sequentially, where $H_{t-1}$ is the previous hidden state followed by the current state, $H_t$. Additionally, the network has Cell states designed similarly to the hidden states. Thus, we have $C_{t-1}\ and\ C_t$ states. The states are time-indexed since we are dealing with time series problems. So, ideally, the current state learns from the previous state. Therefore, there is always state development and improvement. However, this model trains more slowly than GAN models, based on Gated recurrent units (GRUs). Also, the LSTM model is prone to overfitting, and applying the dropout algorithm is complex. This paper provides an efficient way of fitting data and making predictions. However, the LSTM model is fitted as well for comparison's sake.

**Results and Findings**

Descriptive Statistics

| **Statistic** | ***Open*** | ***High*** | ***Low*** | ***Close*** | ***Adj Close*** | ***Volume*** |
|---|---|---|---|---|---|---|
| Mean | 8242.31 | 8301.59 | 8176.99 | 8241.41 | 8241.41 | 103809816.11 |
| Standard Error | 45.60 | 45.76 | 45.42 | 45.60 | 45.60 | 656634.44 |
| Median | 7250.74 | 7313.38 | 7197.66 | 7249.20 | 7249.20 | 94743350.00 |
| Standard Deviation | 3472.04 | 3484.70 | 3458.41 | 3471.91 | 3471.91 | 49999166.42 |
| Kurtosis | -0.94 | -0.94 | -0.93 | -0.94 | -0.94 | 8.18 |
| Skewness | 0.43 | 0.43 | 0.43 | 0.43 | 0.43 | 2.04 |
| Range | 14065.25 | 13970.54 | 14051.76 | 14068.79 | 14068.79 | 510195600.00 |
| Minimum | 2203.97 | 2319.65 | 2188.75 | 2202.96 | 2202.96 | 0.00 |
| Maximum | 16269.22 | 16290.19 | 16240.51 | 16271.75 | 16271.75 | 510195600.00 |
| Count | 5798.00 | 5798.00 | 5798.00 | 5798.00 | 5798.00 | 5798.00 |

*Remark: Summary statistics were computed using the original price data.*

The table above shows the summary statistics of the DAX stock index in question. Among the statistics provided are the mean and standard deviations, for which the mean for opening price is almost equal to the mean of the adjusted and close price, indicating minor differences over time in the stock prices. To fully understand the volatility structure of the model, we report the standard deviation. Interestingly, the opening price also has a higher standard deviation than the close and adjusted closing prices, but the deviation is not significantly different from zero. While we could derive some trends and relationships, we proceeded with our neural nets to understand the deep underlying structure of the stock price movements. A high level and deep understanding of these irregularities is fundamental in this study.

**Historical Index Price Analysis**

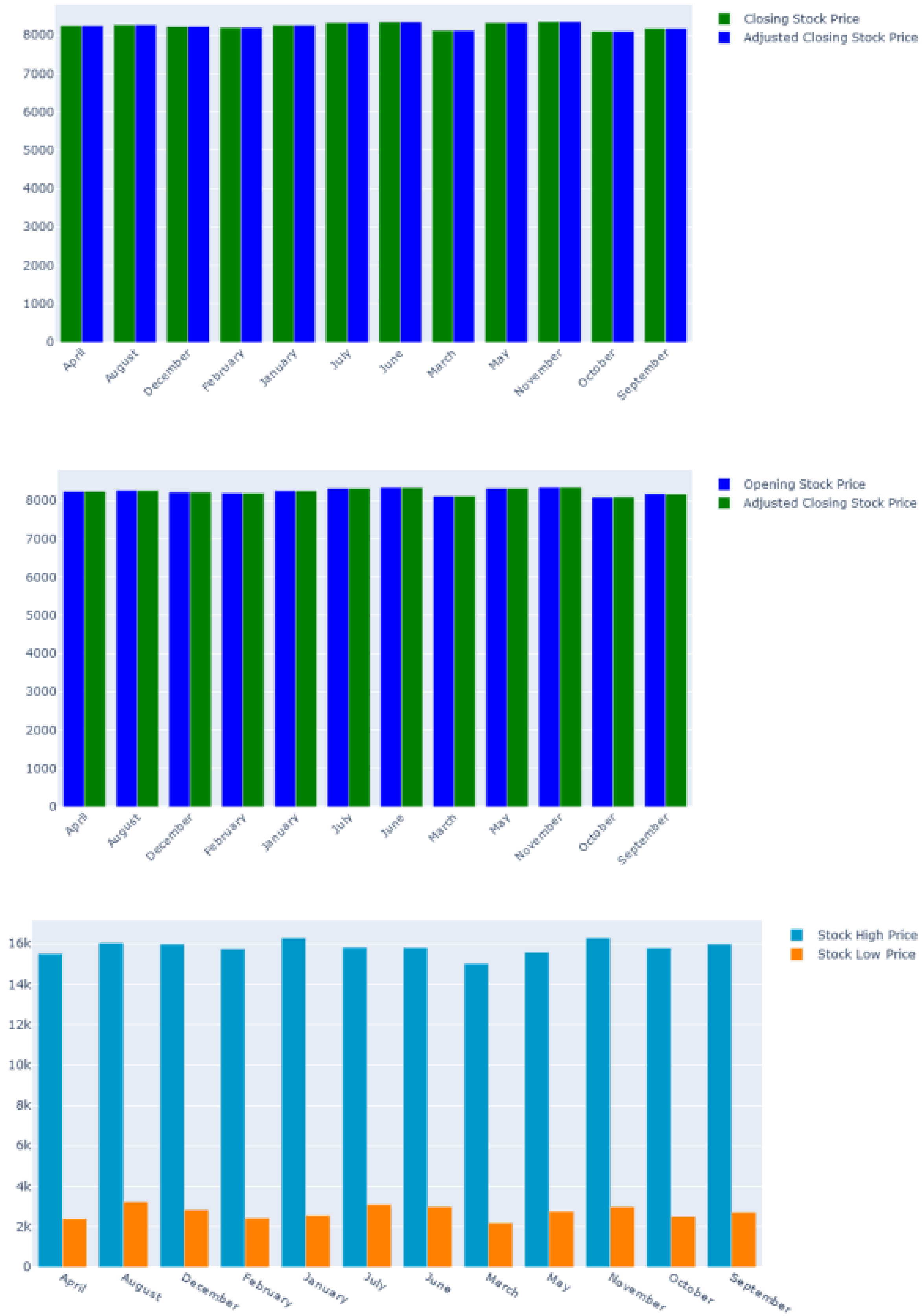

*Figure 3a, b, and c): Month-wise comparison of stock prices.*

We provide charts above comparing the close and open prices against the adjusted closing prices. We notice minor differences across the prices. The close against the adjusted prices is of interest, which indicates slight corporate adjustments such as dividend declarations and stock splits. Figure 3c) compares the high and low stock prices reached over the selected weighted months. Across all the charts, our goal is to identify the month with the highest and lowest prices and the ones with high price variations.

**Combined Stock Price Analysis**

Before implementing, training, and testing our models, we present the combined prices (open, close, high, low, adjusted price) and individual price (close and adjusted close) time series trends.

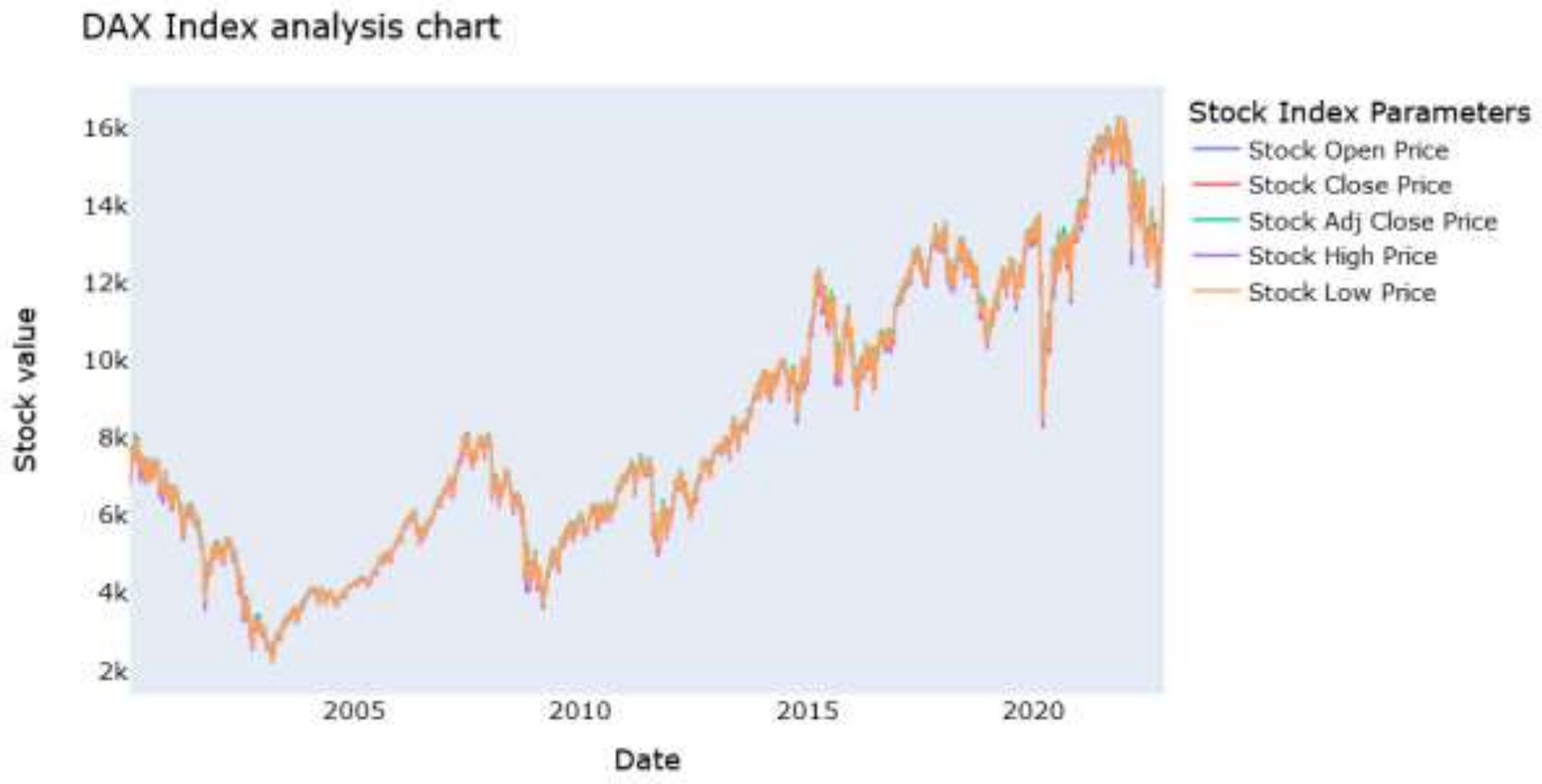

Figure 4a) reports the combined trends among the price metrics recorded, and clearly, the prices follow almost the same trend fashion. We further granularize our analysis by looking closely at the close and adjusted close price.

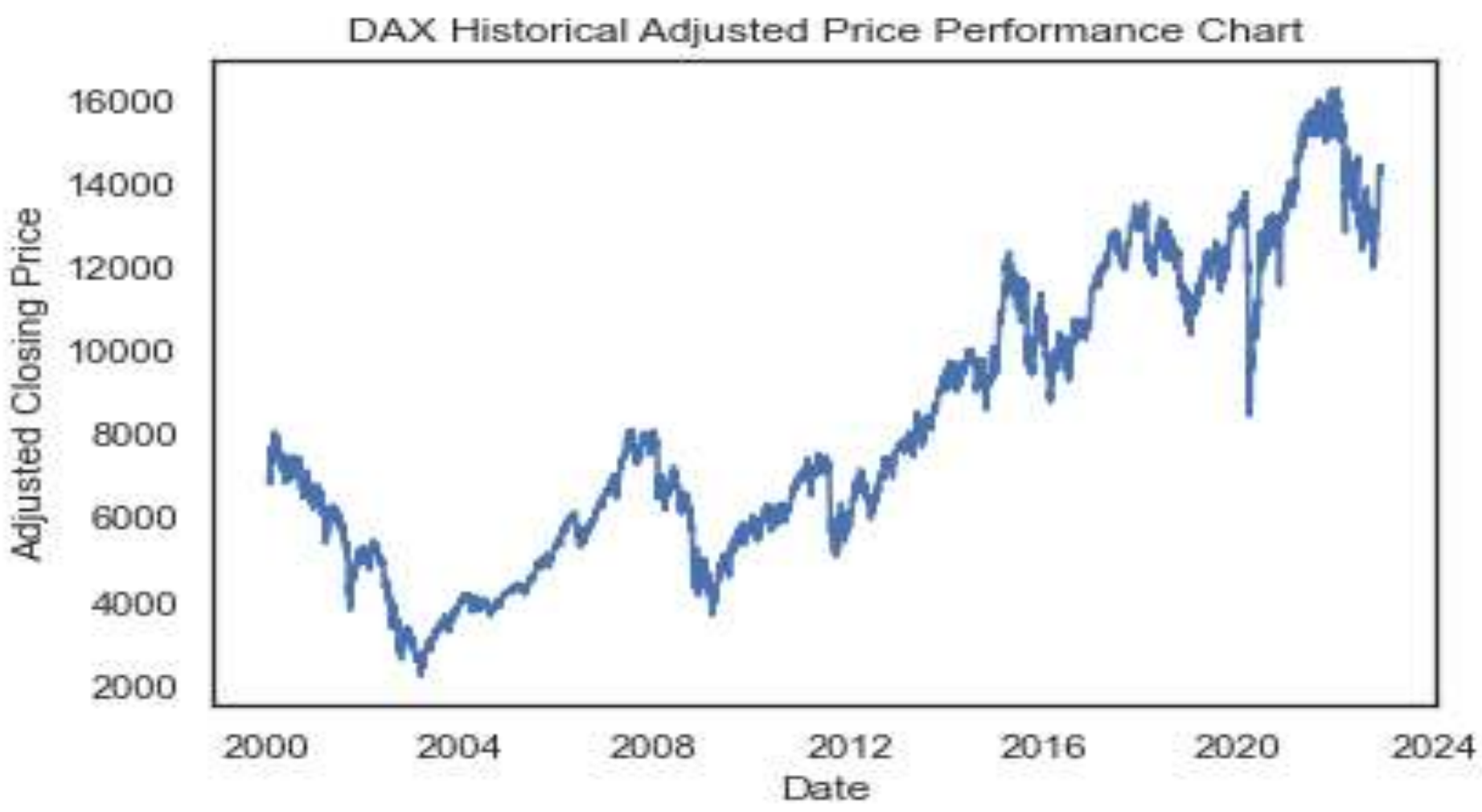

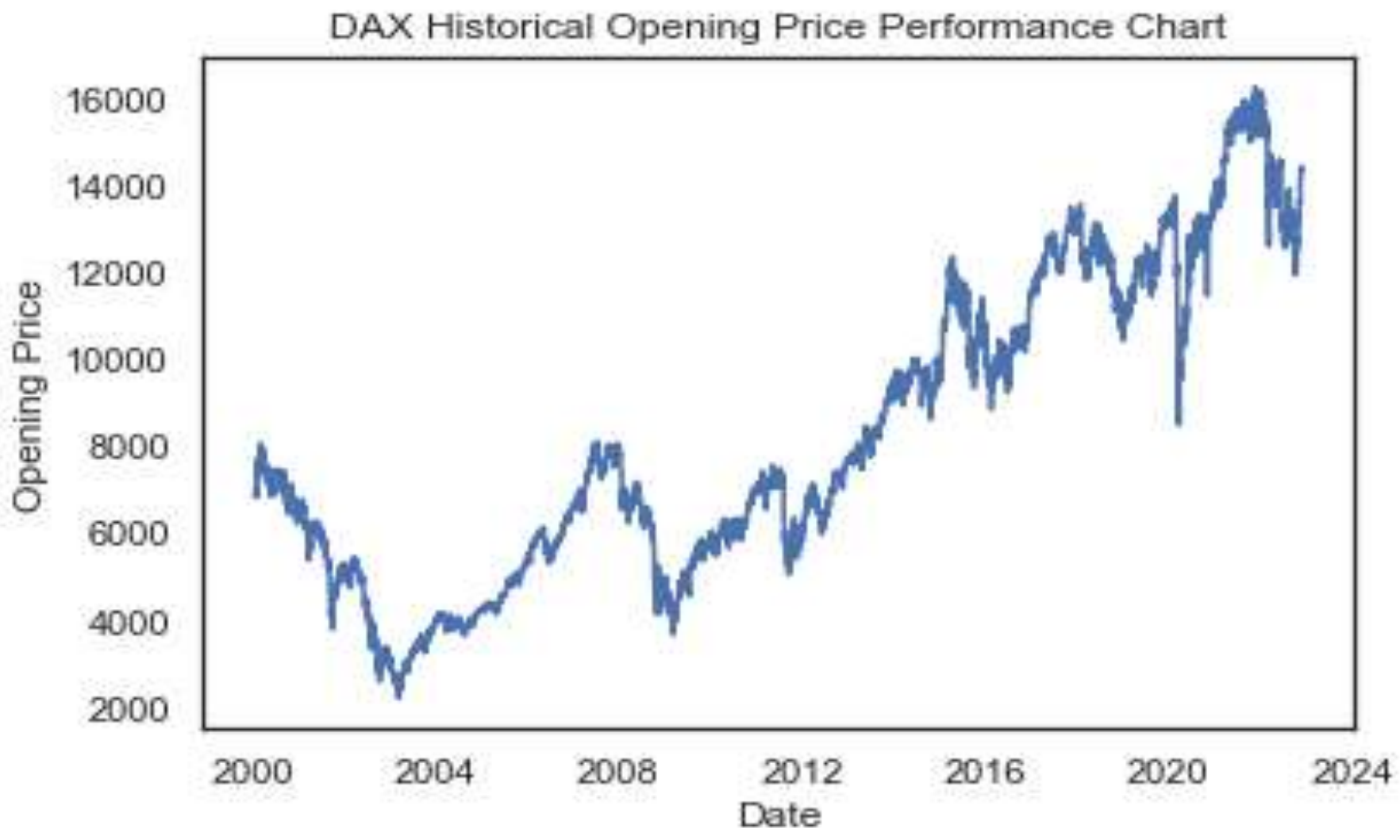

*Figure 4: (b, and c):* From the figures above, we present historical trends of both the opening and adjusted closing prices. On average, we observe rough price paths over the entire period in both opening and closing, with an exceptional downward spike in 2020. We suspect this unusual 'jump' resulted from the COVID-19 shock, which this paper is designed to explore further.

Figure 4d):

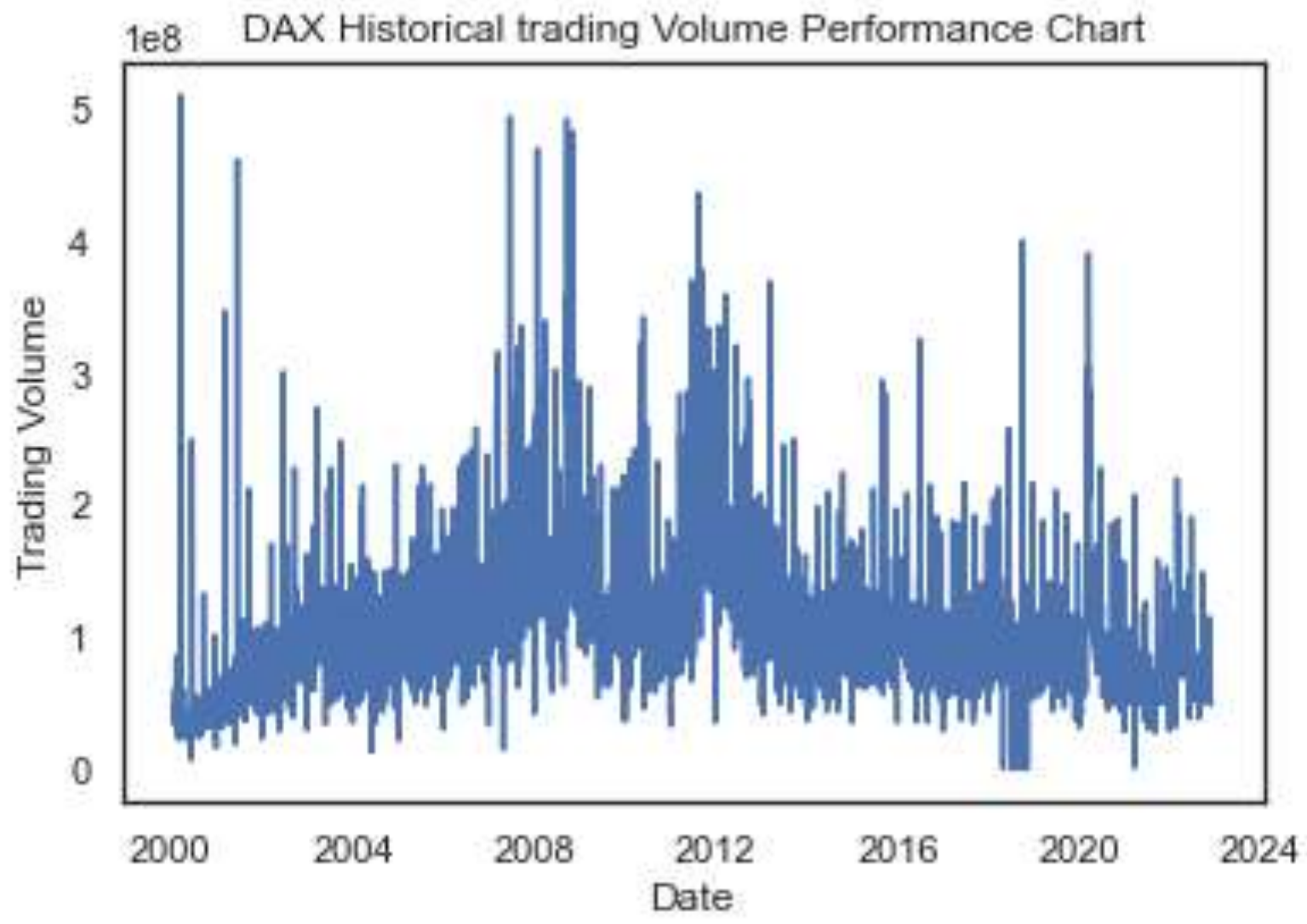

As part of our analysis, we provide a snapshot of the trends associated with trading volumes for the DAX index over six years. On average, the trading volumes are around 2000, with some spikes observable throughout the horizon. These fluctuating trends are associated with risk aversion and investors' lack of investment confidence.

**Correlation Metrics**

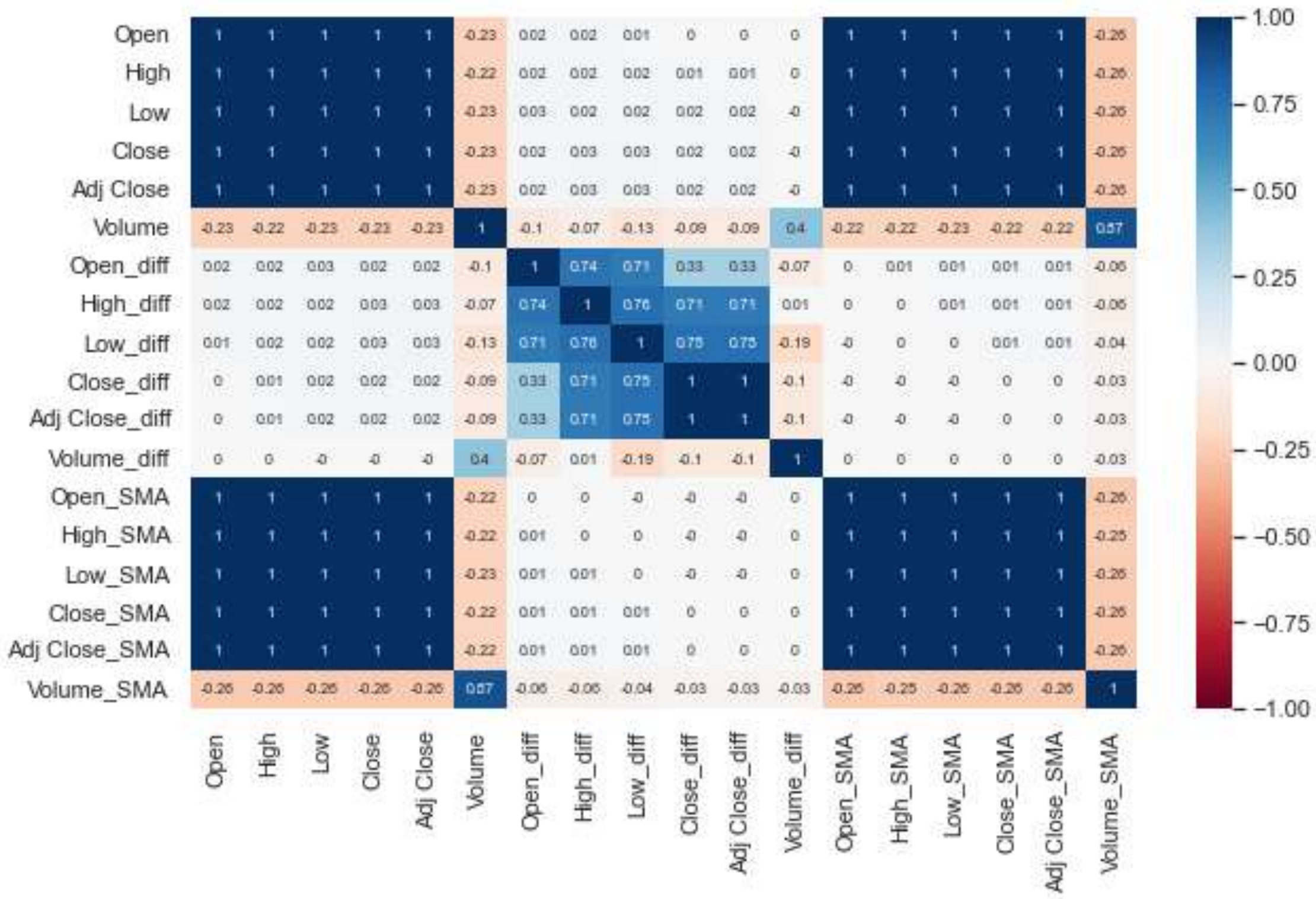

*Table 3: a):* Correlation matrix of the considered variables/indicators. From the cross-correlations among the variables on note, we see that open, high, low, close, and adjusted close prices are highly correlated, with correlation coefficients close to unity (see blue patches). We observe the same correlation clusters among the other new clusters (price differences and simple moving averages prices). Price differences cluster is weakly correlated with the original price variables and the SMA price variables. See whitish patches.

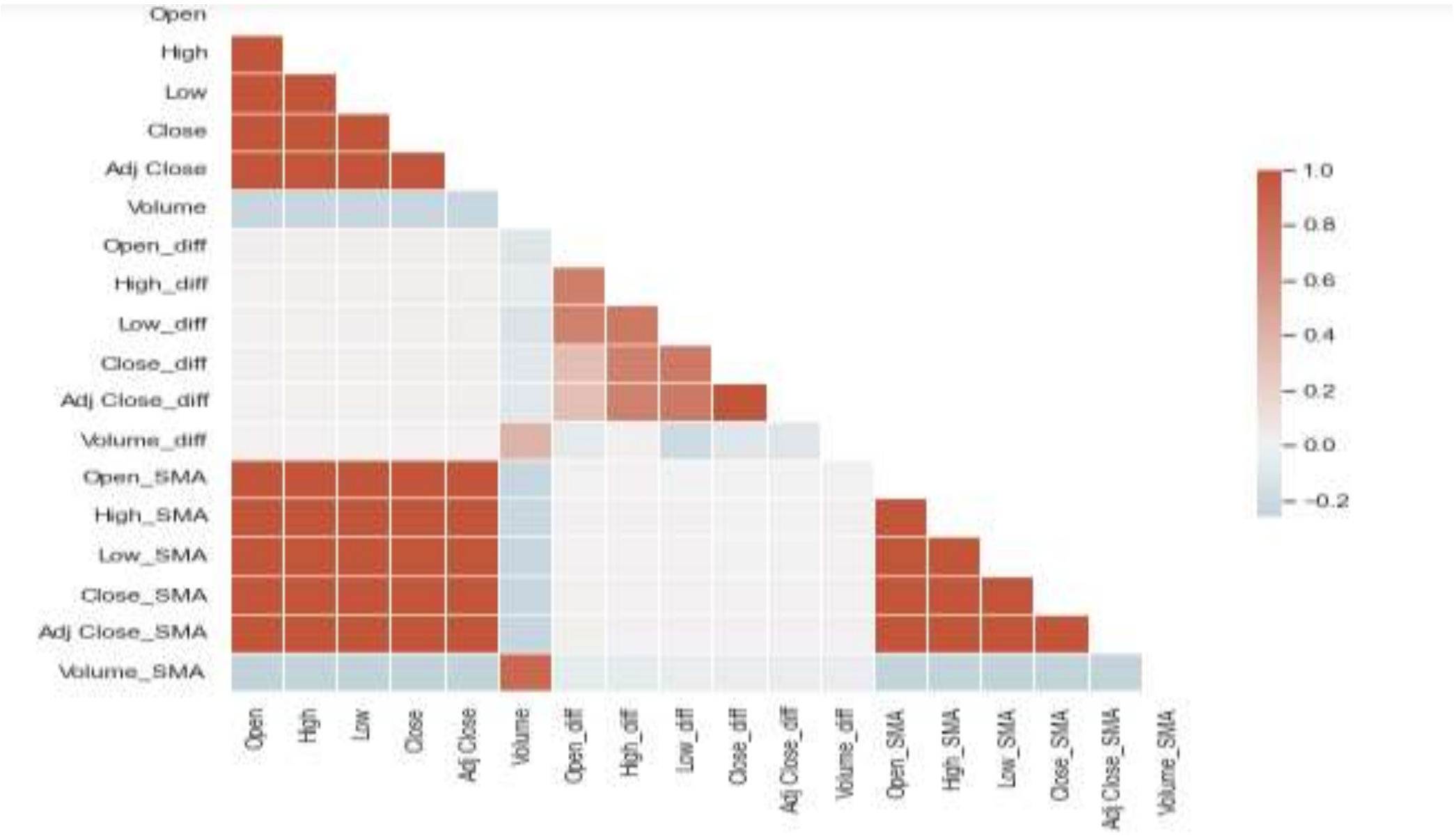

*Table 3b* indicates the same correlation movements in a diagonal format from where the correlation clusters correspond to the correlation hierarchies given below.

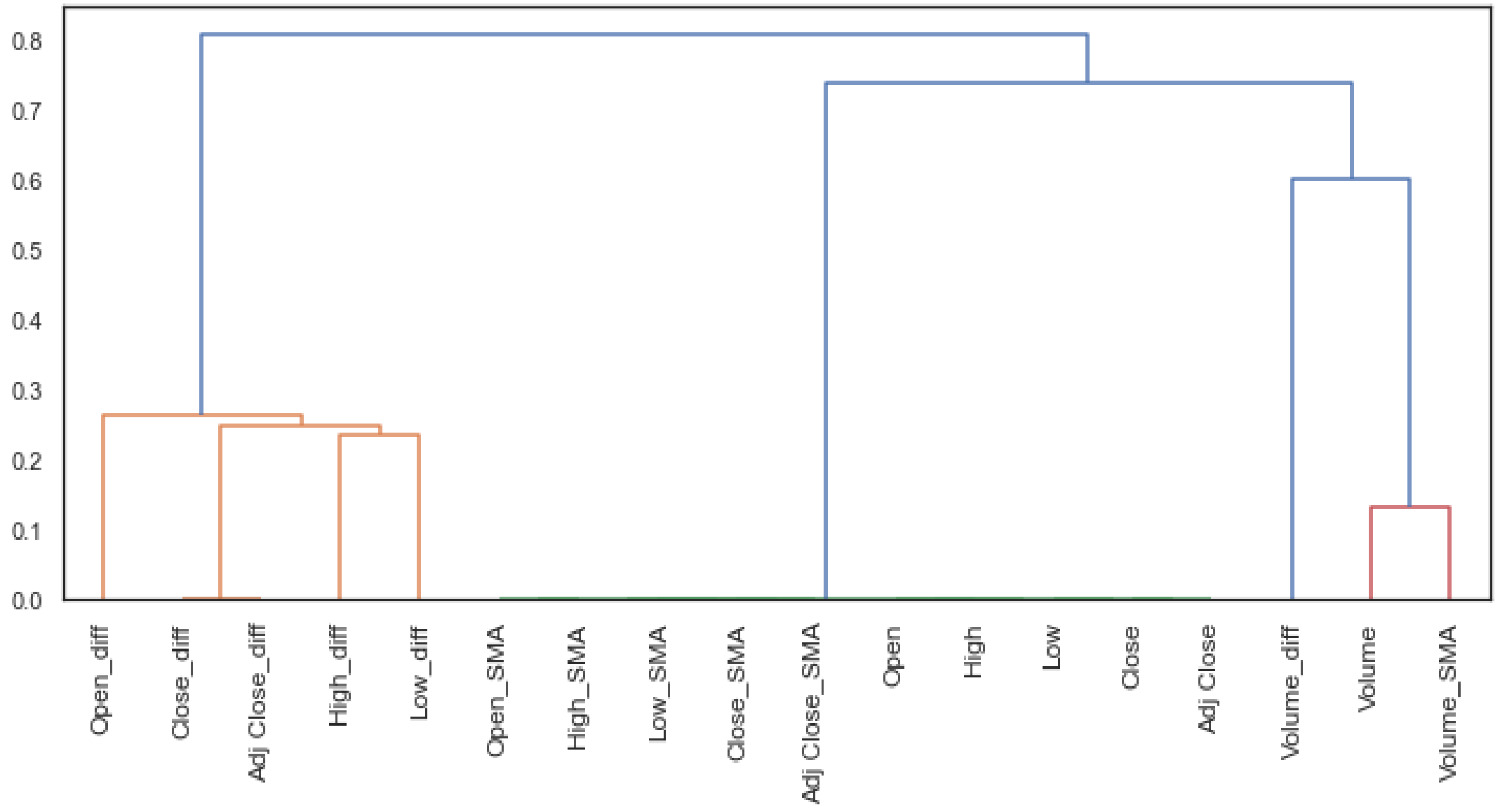

We have roughly two main clusters (red and blue), which tally with the correlation clusters derived in the above matrices. To better understand the correlations and price movements, we provide the following correlation matrix numbers.

| Variable/ Variable | **1** | **2** | **3** | **4** | **5** | **6** | **7** | **8** | **9** | **10** | **11** | **12** |
|---|---|---|---|---|---|---|---|---|---|---|---|---|
| **Open** | 1.00 | | | | | | | | | | | |
| **High** | 1.00 | 1.00 | | | | | | | | | | |
| **Low** | 1.00 | 1.00 | 1.00 | | | | | | | | | |
| **Close** | 1.00 | 1.00 | 1.00 | 1.00 | | | | | | | | |
| **Adj Close** | 1.00 | 1.00 | 1.00 | 1.00 | 1.00 | | | | | | | |
| **Volume** | 0.23 | 0.22 | 0.23 | 0.23 | 0.23 | 1.00 | | | | | | |
| **Open_diff** | 0.02 | 0.02 | 0.03 | 0.02 | 0.02 | 0.10 | 1.00 | | | | | |
| **High_diff** | 0.02 | 0.02 | 0.02 | 0.03 | 0.03 | 0.07 | 0.74 | 1.00 | | | | |
| **Low_diff** | 0.01 | 0.02 | 0.02 | 0.03 | 0.03 | 0.13 | 0.71 | 0.76 | 1.00 | | | |
| **Close_diff** | 0.00 | 0.01 | 0.02 | 0.02 | 0.02 | 0.09 | 0.33 | 0.71 | 0.75 | 1.00 | | |
| **Adj_Close_diff** | 0.00 | 0.01 | 0.02 | 0.02 | 0.02 | 0.09 | 0.33 | 0.71 | 0.75 | 1.00 | | |
| **Volume_diff** | 0.00 | 0.00 | 0.00 | 0.00 | 0.00 | 0.40 | 0.07 | 0.01 | 0.19 | 0.10 | 1.00 | |

Table 4: Correlation matrix across our indicators. High coefficient values correspond to the blue patches in the matrix and the hierarchy above, and small values are associated with the whitish patches, which consistently agrees with the summary statistics and some descriptive trend changes provided earlier.

**Main Results**

In this section, we provide a detailed presentation and explanation of each of the results from our models.

GRU Model

Ideally, the ability of the GRU model to store and filter information from the input networks as they update their gates allowed us to fit and forecast the closing prices in a stochastic stock market setup. The idea is that if these GRU models are well-trained, they can perform well, as indicated below.

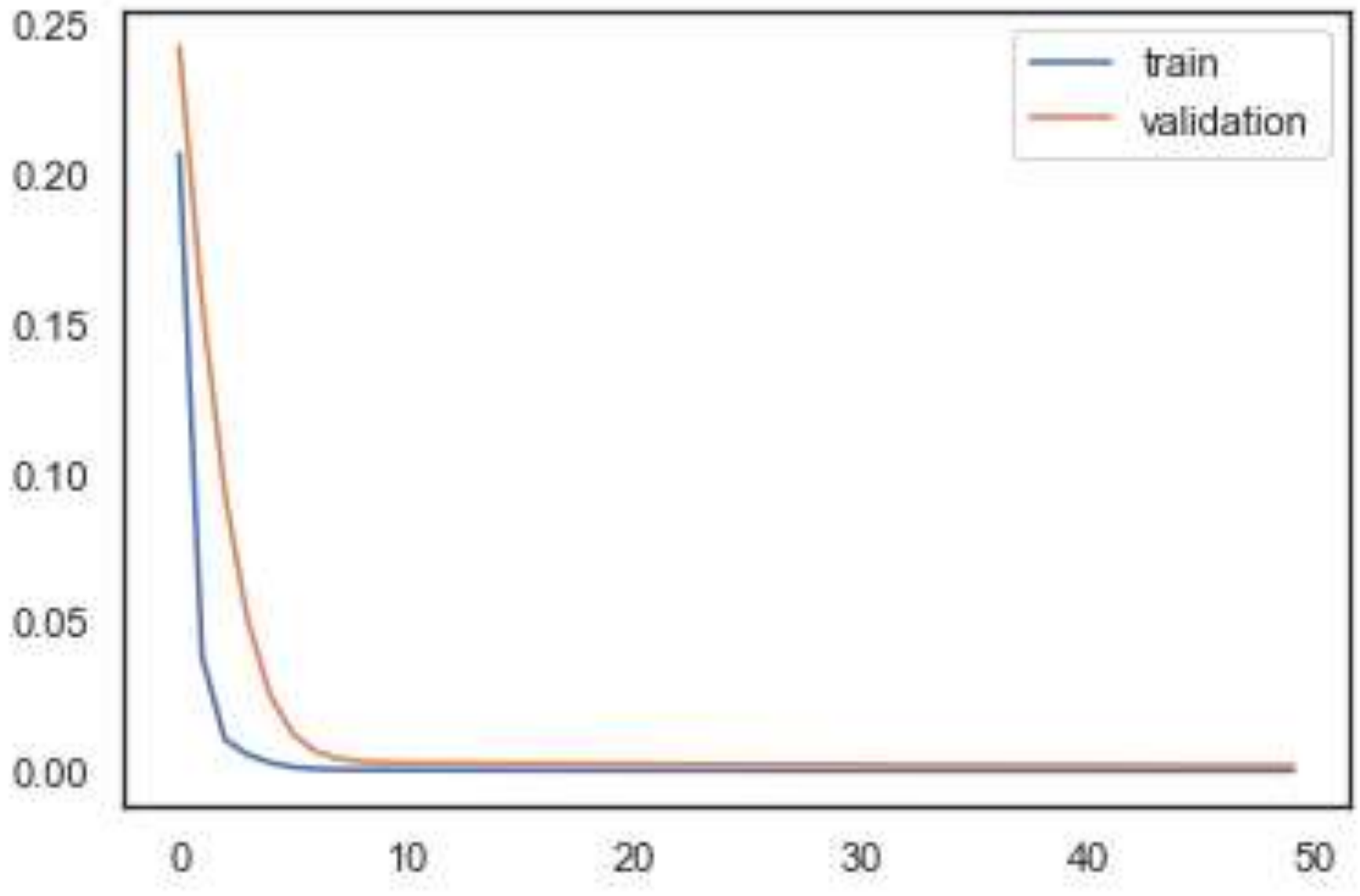

*Figure 5:* First, we split our dataset into two subsets comprising the training and test sets. Our models are trained using the training set and validated/tested using the test set. We employed a ratio of 7:3 for the training and test set, respectively. As such, from the figure above, we present the model's performance on both the training and testing sets. As indicated above, performance error is marginal enough to proceed with the GAN and WGAN training, including forecasting. Below, we provide the training and forecasting results in more depth.

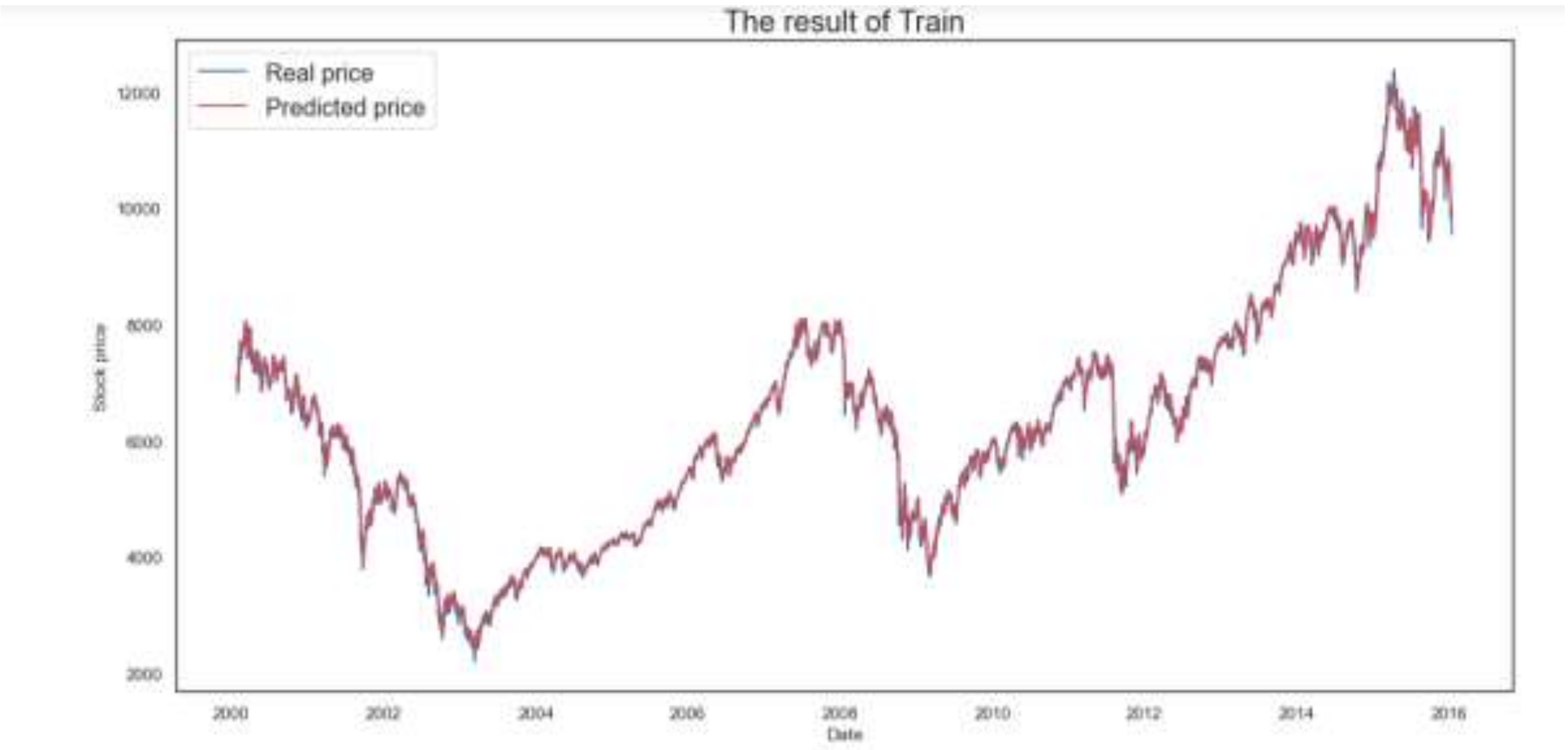

*Figure 6:* The model shows the real and predicted prices using the training set. The model fits the data well, as indicated by the small RMSE. See the validation section. This indicates the robustness of our GRU model. We applied the same network and architecture on the test set, whose results are shown below.

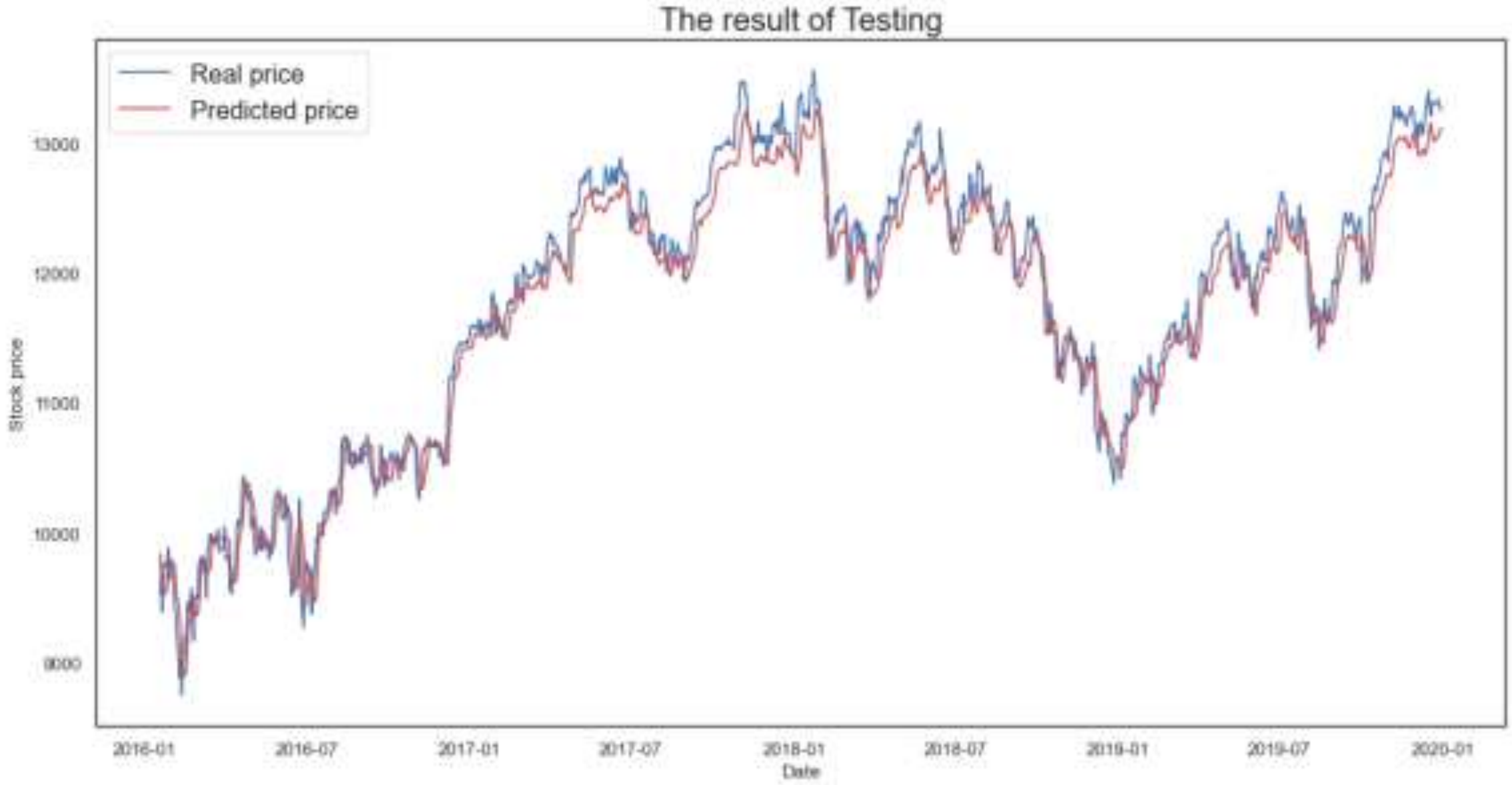

*Figure 7:* Our test set is smaller than the training set, hence the differences in price paths-thickness and roughness. Interestingly, our GRU model could fit and predict the closing prices with low RMSE errors. The real and predicted prices imitate almost the same pattern over space.

TimeGANS Model

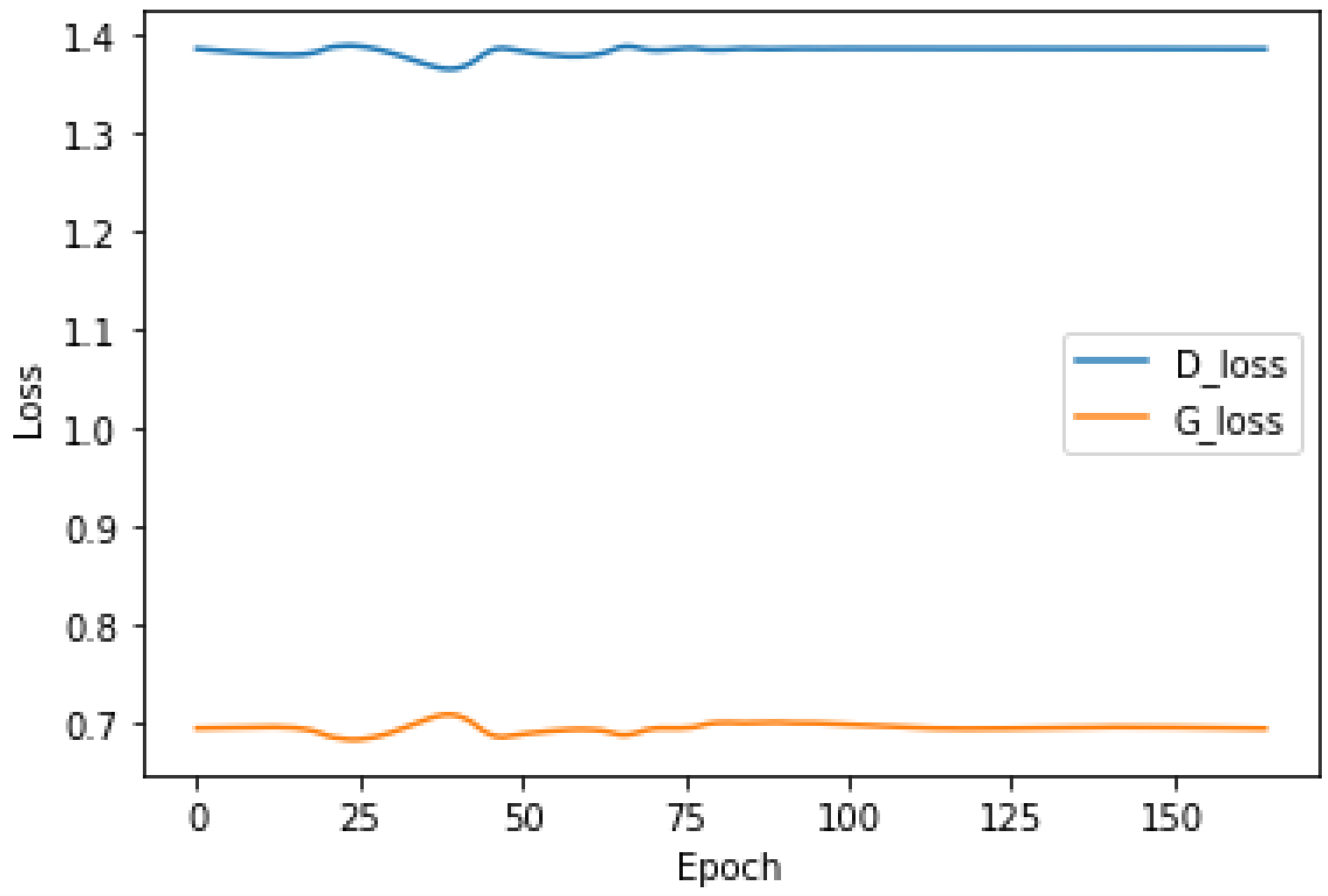

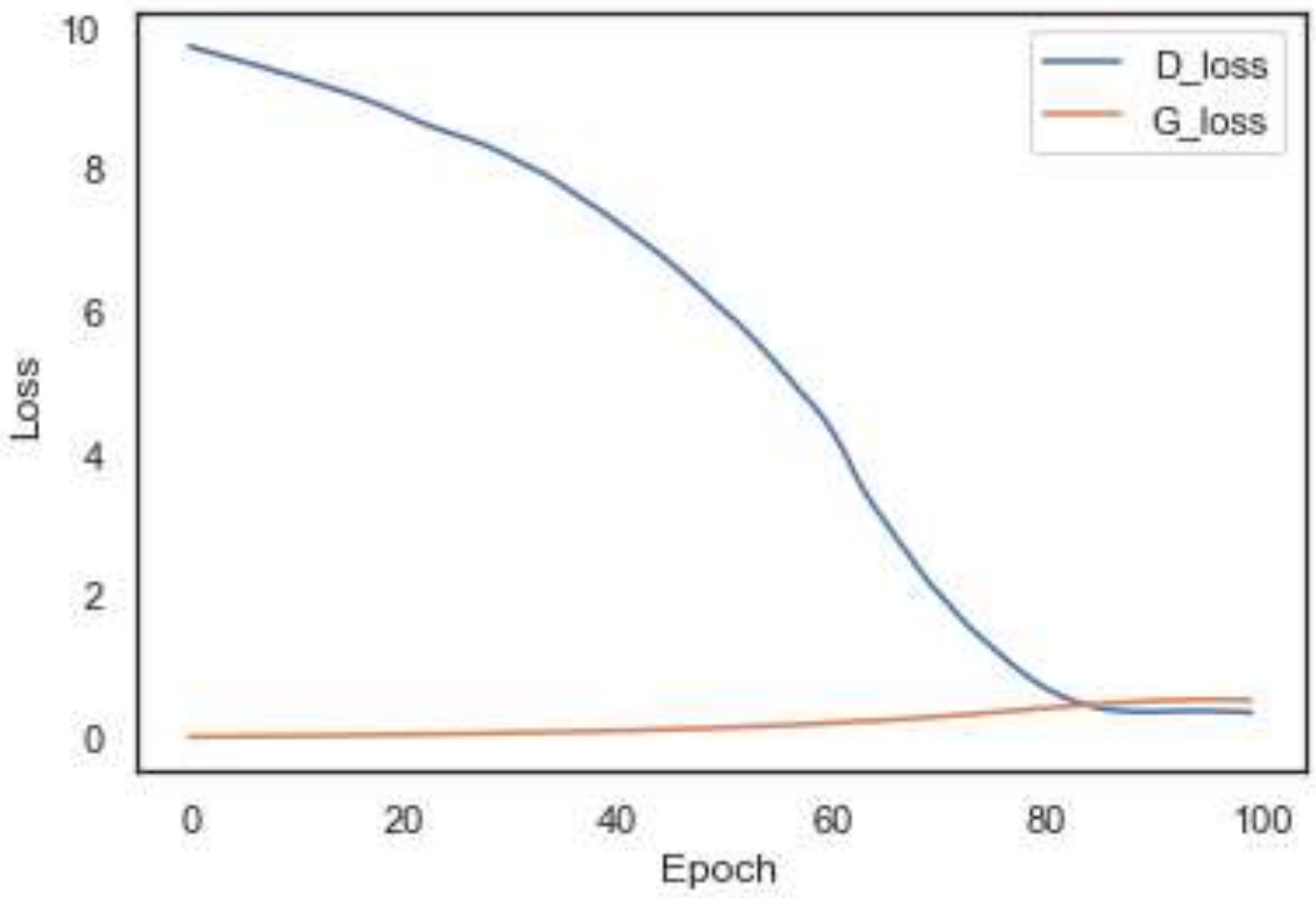

*The above schemes show the loss function results between the generator and the discriminator, where*

*Figure 8:* Unlike the GRU, where the model is fitted and compared on the train and test sets, GANS are based on the discriminator and generator in a game setup. As described earlier, the generator attempts to generate new observations-in, in this case, closing prices, and the discriminator attempts to classify the newly generated data as real or fake. The fake data is regenerated until the discriminator fails to differentiate the data from the real ones. To fully figure out the game setup, we present the losses from both the discriminator and generator, and the discriminator retains higher losses than their generator counterparts.

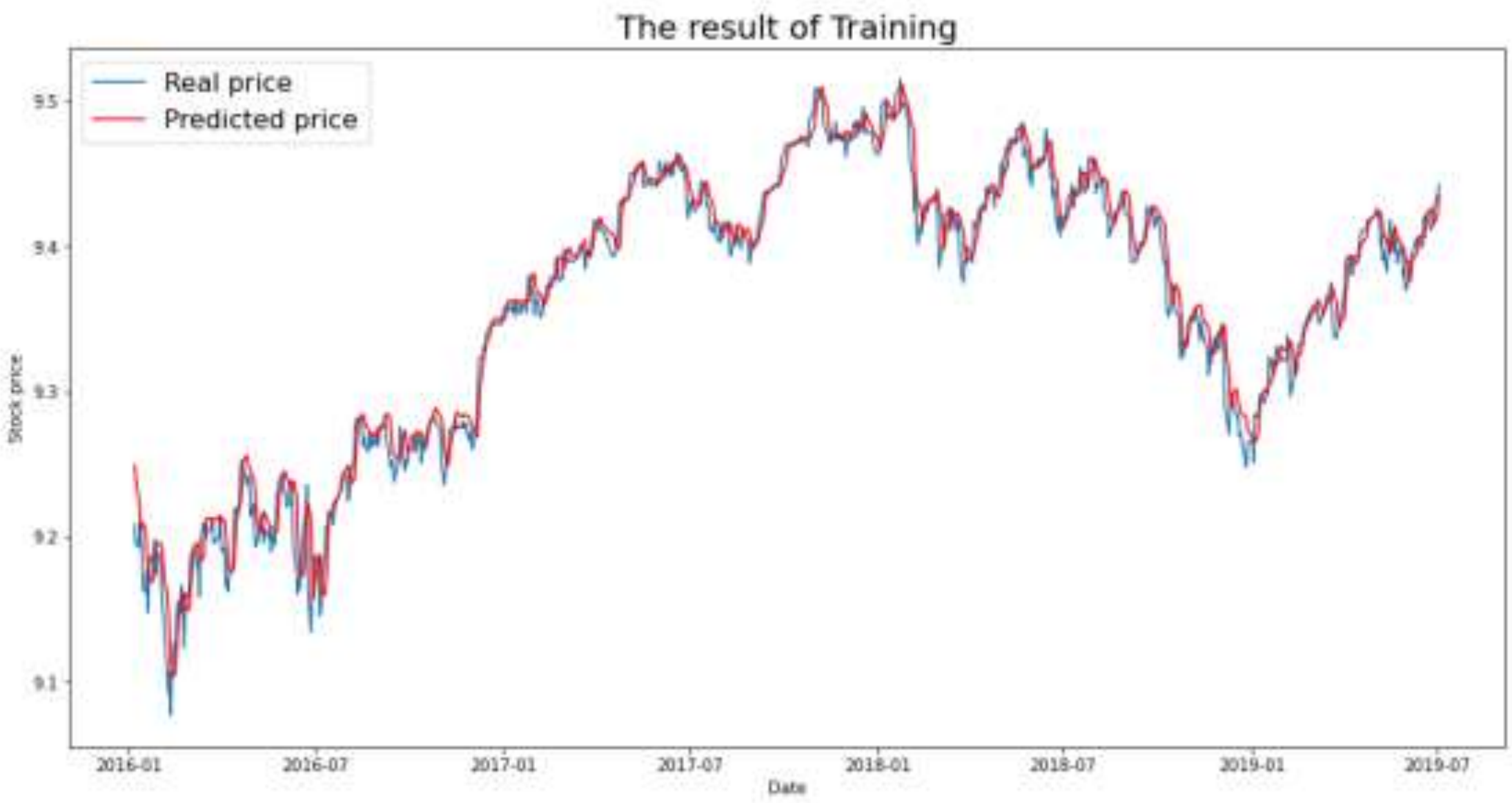

*Figure 9:* Following training our GAN model multiple times until the losses are small, the model produced the graph above, comparing the real and predicted closing prices. The results support the GRU model for fitting and predicting closing stock prices under structural shocks. Further, we fitted the GAN model on the stock returns. We employed the same architecture, except for the parameters, which we tuned to match the case of stock returns. See below.

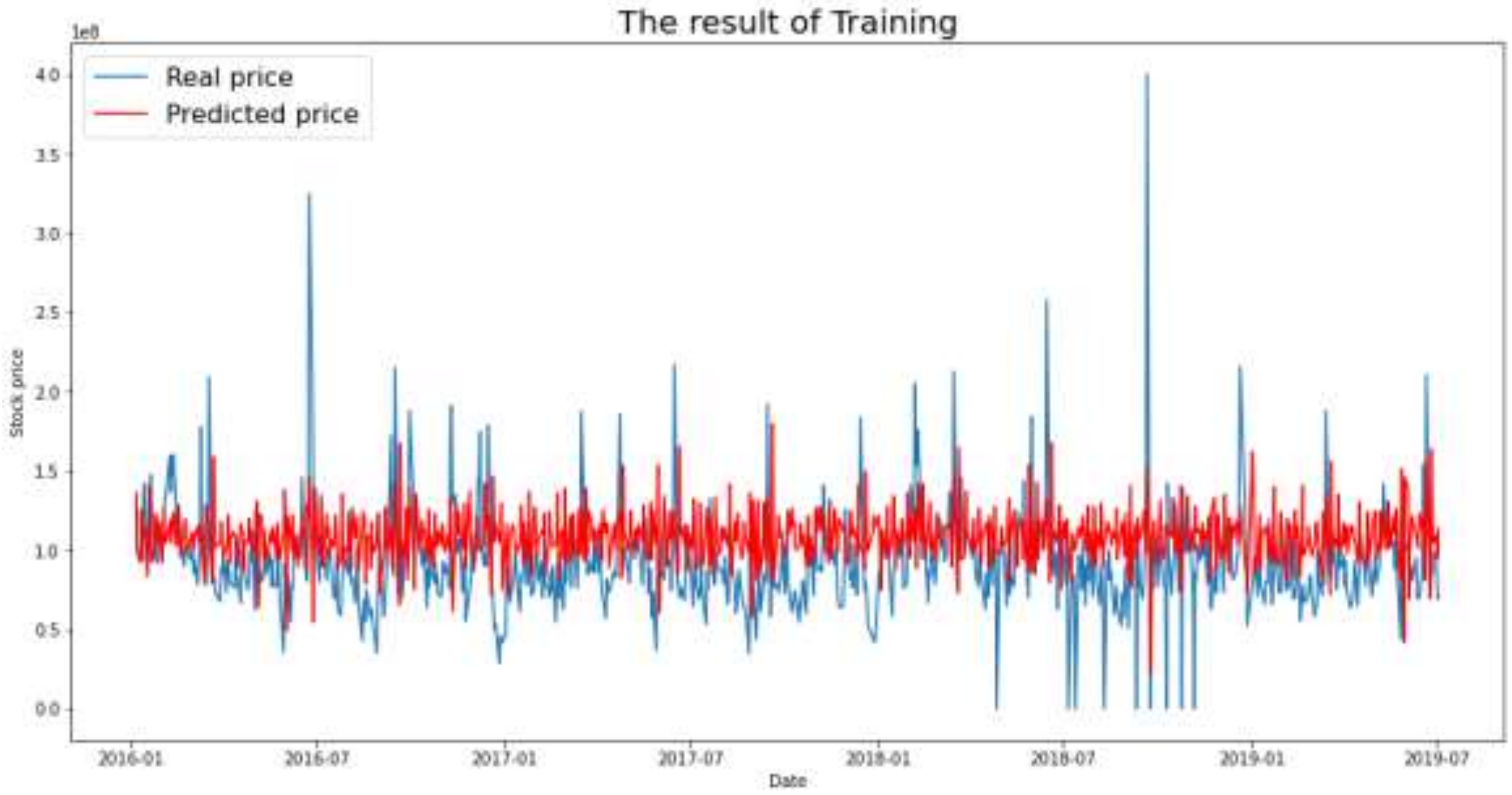

*Figure 10: Stock returns based on closing prices.* The model fitted the training set of the returns (real returns well). However, the predicted model failed to provide predicted returns with minimum errors.

WGANs Model

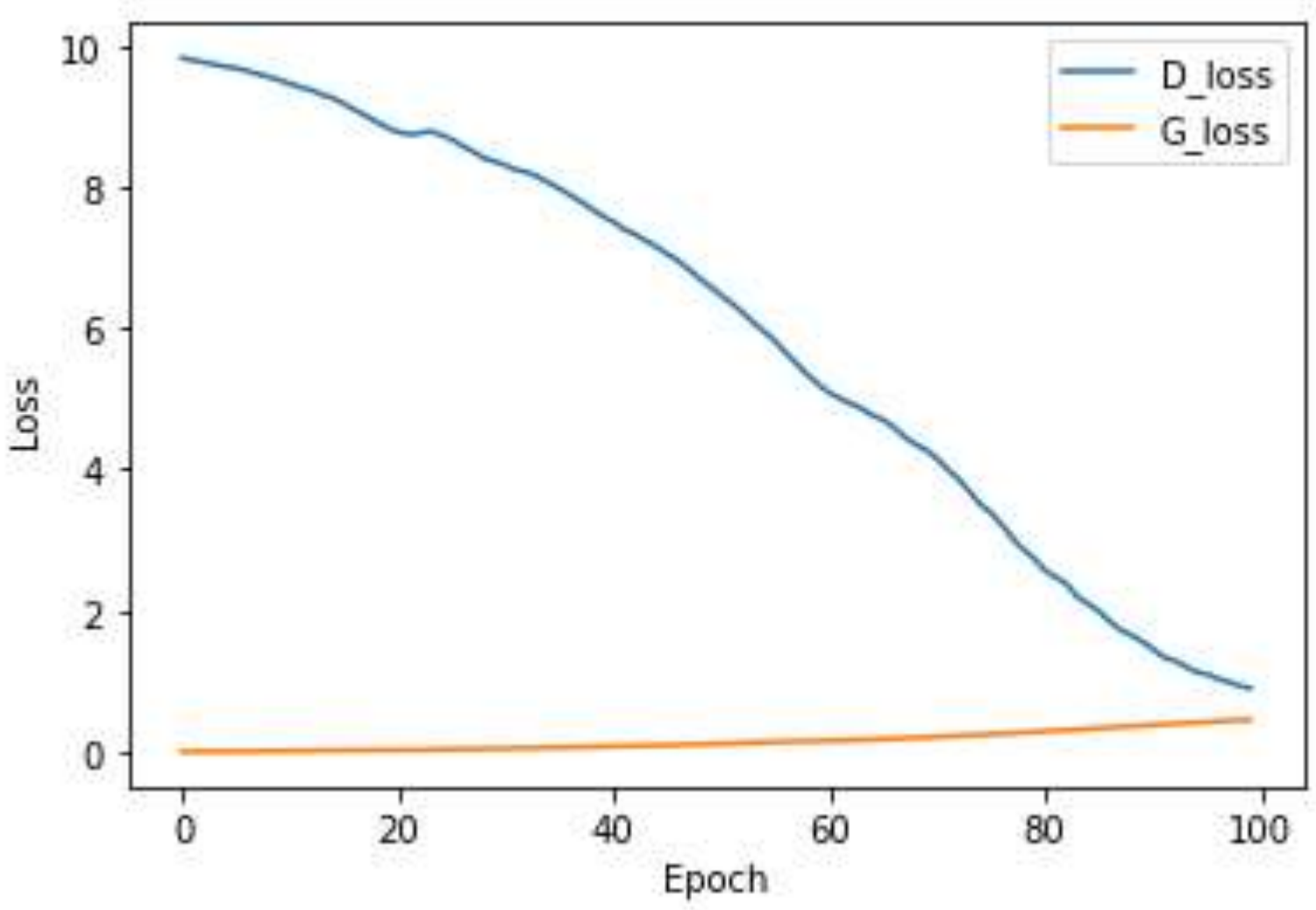

*Figure 11:* WGANs are a special type of the GAN network fitted in the previous section. It differs from the GAN model in that it uses the loss function based on the Wasserstein distance metric, while our GAN above and other models are based on the Euclidean metrics. Also, its loss function gives a termination criterion for model evaluation. Like in the GAN architecture, we provide the behavior of the loss functions of the competing generator and discriminator. Interestingly, we noted that the two functions tend to converge as we increase the number of training epochs (instances). We also provide the fitted model results below.

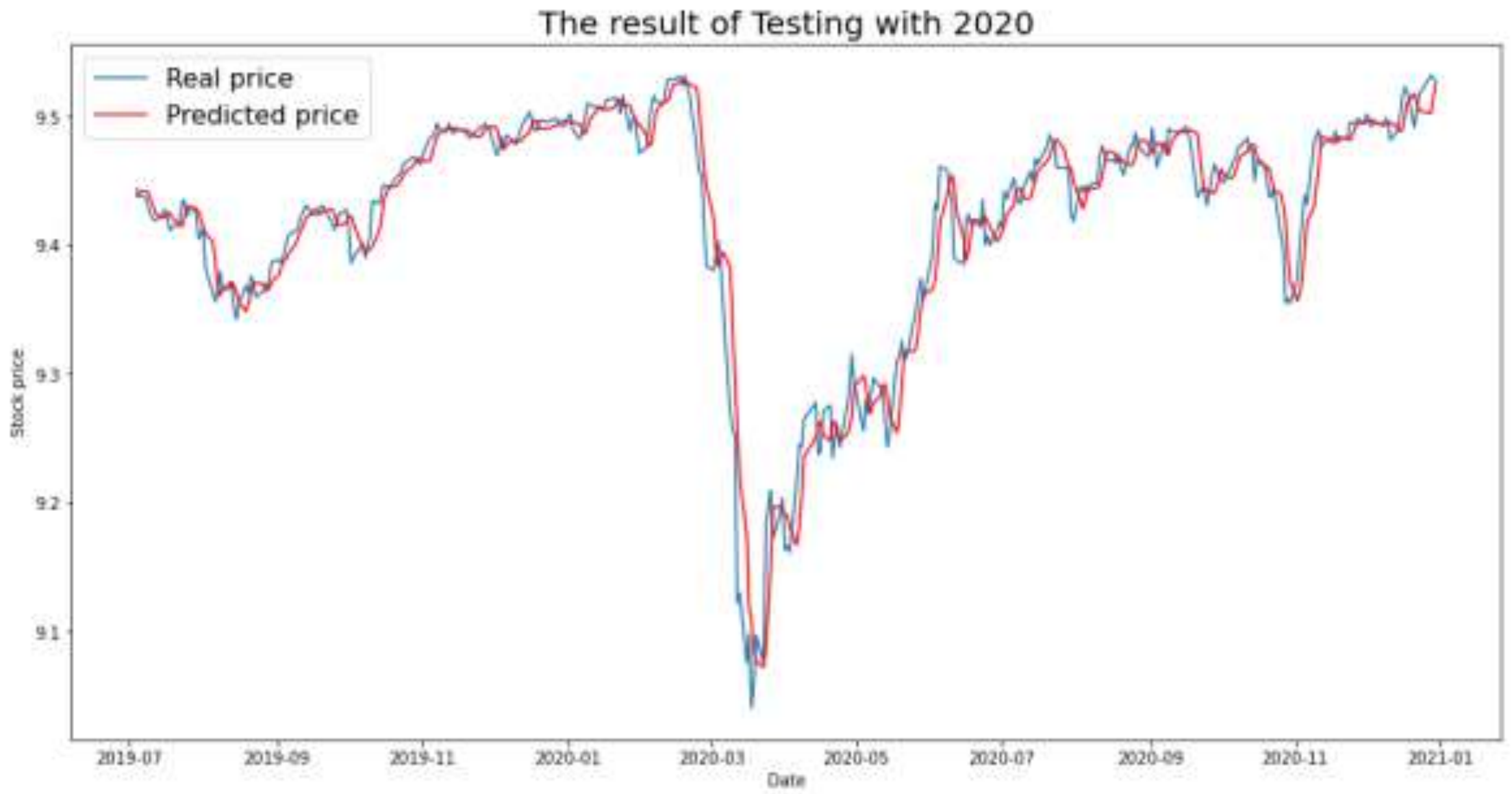

*Figure 12*: The WGAN model generated the new data instances and classified them as real or predicted. We then fitted and presented the resulting closing prices, which were indifferent in our case. This is consistent with the convergence idea given above on the loss functions.

LSTM Model

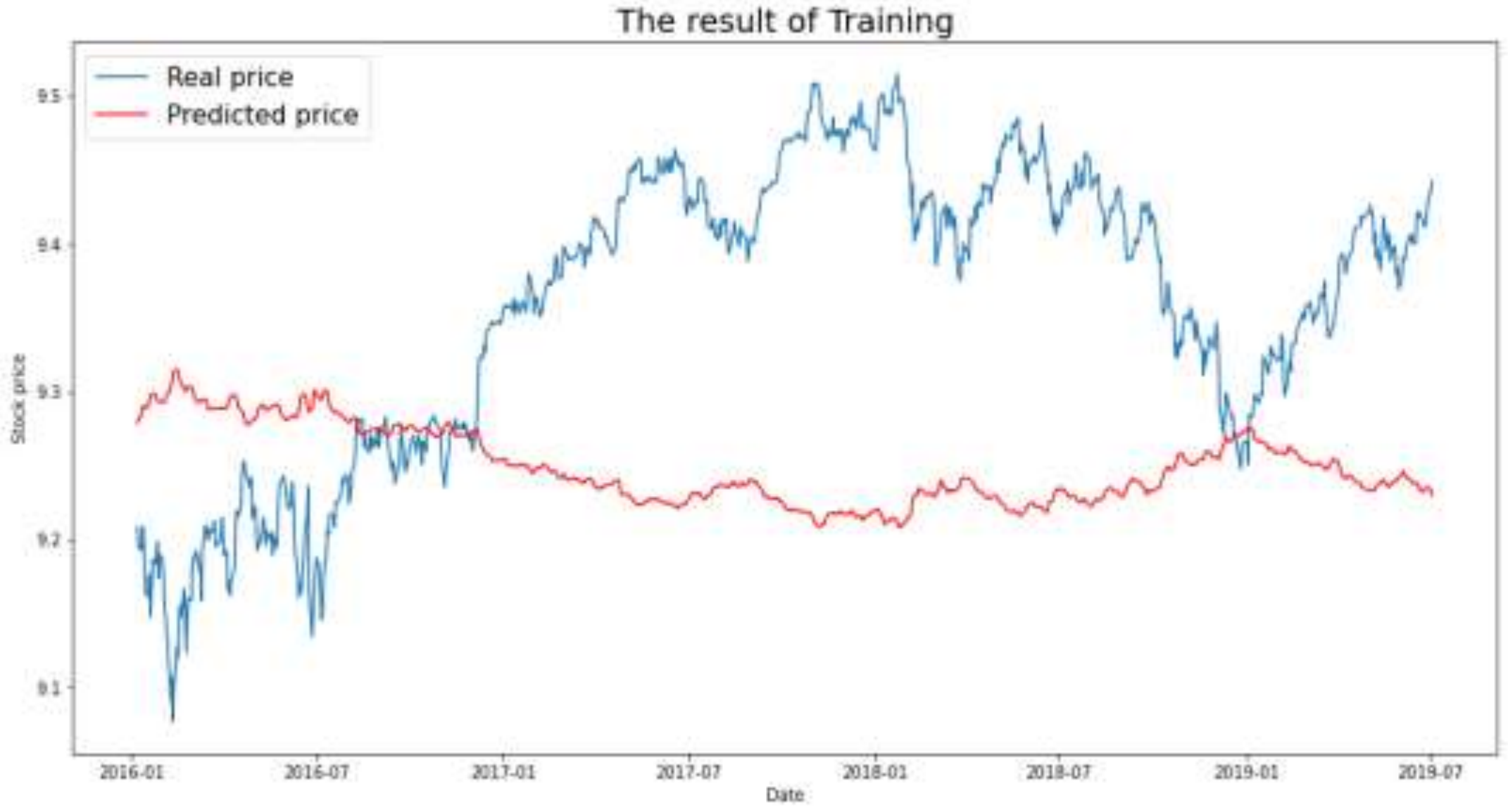

*Figure 13:* The LSTM model fits the real and predicted prices as above. One of the weaknesses of the LSTM model is its proneness to overfitting and its sensitivity to different random weight initializations. Also, the LSTM requires more datasets for plausible results, and they usually could perform better on small to medium datasets like ours. This points out the power and robustness of the GANS and WGANS networks on fine modeling and time series forecasting in small financial data sets. As in the figure below on the test set, the LSTM model fails to fit and accurately predict the closing prices.

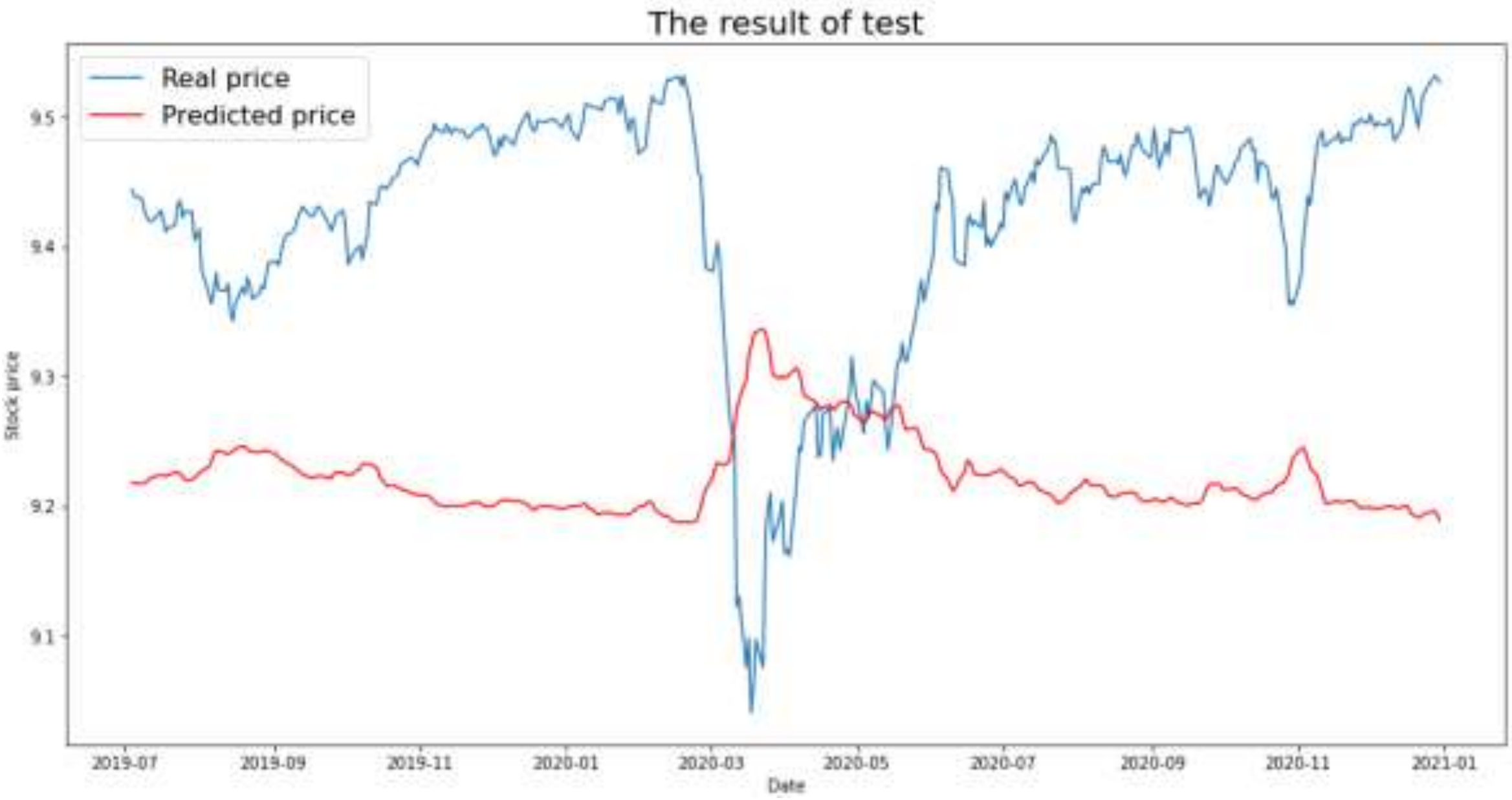

We attempted to predict the closing prices on the test set, and the model provided a deviated fit, which is extensively featured by high forecasting RMSE errors as we elaborate subsequently.

**Model Validation and cross-comparison across prediction horizons.**

| MODEL | RMSE | MAPE | Number of Hidden Layers | EPOCH Number |
|---|---|---|---|---|
| WGAN | 0.553 | 0.515 | 6 | 250 |
| LSTM | 0.623 | 0.662 | 6 | 150 |
| TimeGAN | 0.347 | 0.3804 | 6 | 250 |
| GRU | 0.3901 | 0.4073 | 6 | 50 |

We report the weighted average root mean square errors and the mean absolute percentage errors obtained from each model trained and fitted in this study. Our results suggest that the LSTM model retained high error values with TimeGAN, GRU, and WGAN provided fits with small values. We compared our models by perturbing the number of layers and epochs and noting each error accumulation for each model over the 10, 40, and 80-day time horizons. Interestingly, the errors decrease with an increase in the number of training epochs and hidden layers. This means the Generative models do well under multiple training hidden and dense layers with multiple epochs.

**Discussions**

Forecasting the closing prices of stocks is challenging in stock markets due to the non-linearity of the market, data unavailability, and market volatility, among other factors. At the same time, making decisions based on market forecasts and predictions is challenging, especially in the presence of structural shocks like COVID-19 and the Russia-Ukraine war. This paper attempts to provide versatile and more robust methods of forecasting closing prices in stock markets in the presence of structural shocks, particularly COVID-19. Generative adversarial networks of a time series nature were used. In particular, we fitted the GRU model, WGAN, TimeGAN (primary model), and LSTM (benchmark). While the LSTM was used to validate our WGAN and TimeGAN models, GRU was employed as the base model for our approach. Using daily data for the DAX index from 2016 to 2022, our results suggested that stock prices can be better forecasted and predicated using the TimeGAN and WGAN models than the LSTM model, as in Table 5 (forecasting errors). Also, our models were able to capture the underlying and hidden market irregularities, and price jumps in stock prices in times of COVID-19 shock. This marks the extension of the work initiated by Staffini (2022). The plausibility of our results lies in their usefulness to investors when making decisions under rough and uncertain market conditions. In addition, our TimeGAN networks are valuable when faced with small financial datasets where patterns are challenging to extract. Nevertheless, hyper-parameter tuning remains a problem when training these deep learning models, and there remains an avenue for future researchers to provide a more stable and analytical way of parameter tuning.